\documentclass[acmsmall,screen,authorversion,nonacm]{acmart}
\usepackage{wrapfig}
\usepackage{rotating}
\usepackage{floatflt}
\usepackage{graphicx}
\usepackage{subfig}
\usepackage{fancyhdr}
\AtBeginDocument{%
  \providecommand\BibTeX{{%
    \normalfont B\kern-0.5em{\scshape i\kern-0.25em b}\kern-0.8em\TeX}}}

\AtBeginDocument{%
    \addtolength{\footskip}{2.0\baselineskip}%
    \fancyfoot[L]{\textit{\textbf{Preprint - Submitted for peer review}}}%
}

\setcopyright{acmcopyright}
\copyrightyear{2018}
\acmYear{2018}
\acmDOI{10.1145/1122445.1122456}

\acmConference[Woodstock '18]{Woodstock '18: ACM Symposium on Neural
  Gaze Detection}{June 03--05, 2018}{Woodstock, NY}
\acmBooktitle{Woodstock '18: ACM Symposium on Neural Gaze Detection,
  June 03--05, 2018, Woodstock, NY}
\acmPrice{15.00}
\acmISBN{978-1-4503-XXXX-X/18/06}

\begin{document}

\title[ AR Supported Collaboration for Emergency Response]{\system: Augmented Reality Supported Collaboration for UAV Driven Emergency Response Systems}

\author{Ankit Agrawal}
\email{aagrawa2@nd.edu}
\author{Jane Cleland-Huang}
\email{JaneHuange@nd.edu}
\affiliation{
  \institution{University of Notre Dame}
  \city{Notre Dame}
  \state{Indiana}
  \country{USA}
  }

  

\renewcommand{\shortauthors}{Ankit Agrawal and Jane Cleland-Huang}

\newcommand{\framework}{ERC Framework} 

\newcommand{\droneresponse}{ResponseTeam }

\newcommand{\dronear}{Flying-AR }
\newcommand{\Dronear}{Flying-AR}
\newcommand{\dronearx}{Flying-AR} 

\newcommand{\humanar}{FirstResponse-AR }
\newcommand{\Humanar}{FirstResponse-AR}

\newcommand{\missionplan}{Mission-Control }
\newcommand{\Missionplan}{Mission-Control}

\newcommand{\system}{RescueAR }
\newcommand{\systemx}{RescueAR}
\newcommand{\System}{RescueAR}

\newcommand{\jch}[1]{{\textcolor{purple}{#1}}}
\newcommand{\ankit}[1]{{\textcolor{blue}{#1}}}

\begin{abstract}
Emergency response events are fast-paced, noisy, and they require teamwork to accomplish the mission. Furthermore, the increasing deployment of Unmanned Aerial Vehicles (UAVs) alongside emergency responders, demands a new form of partnership between humans and UAVs. Traditional radio-based information exchange between humans during an emergency response suffers from a lack of visualization and often results in miscommunication. This paper presents a novel collaboration platform: \systemx, which utilizes the paradigm of Location-based Augmented Reality to geotag, share, and visualize information. \system aims to support the two-way communication between humans and UAVs, facilitate collaboration across diverse responders, and visualize scene information relevant to the rescue team's role. According to our feasibility study, user study, followed by a focus group session with police officers, \system can support rescue teams in developing the spatial cognition of the scene, facilitate the exchange of geolocation information, and complement existing communication tools during the UAV-supported emergency response.

\end{abstract}



\begin{teaserfigure}
  \includegraphics[width=\textwidth]{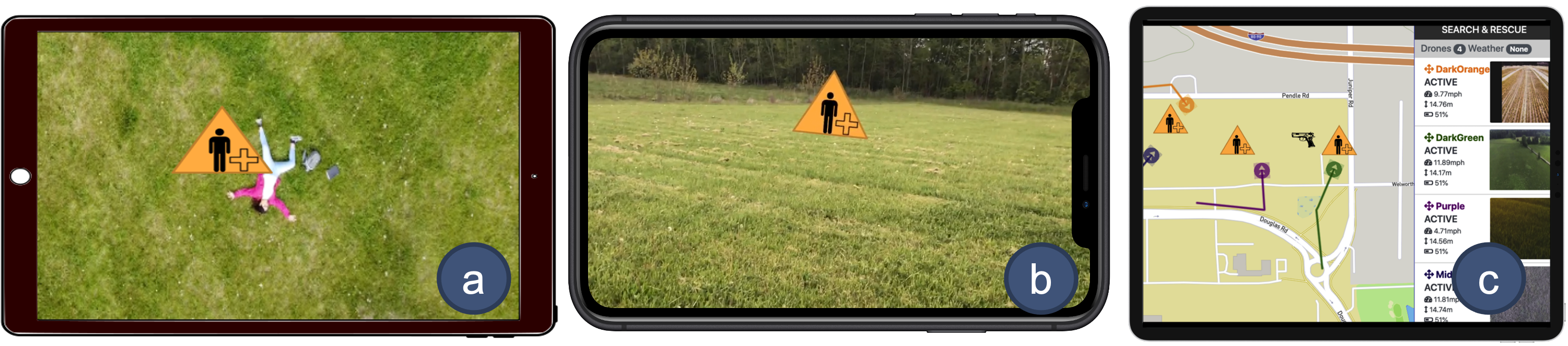}
  \caption{\system utilizes location-based augmented reality technology to facilitate the exchange of information among rescue team members during a UAV-driven emergency response. (a) \dronear: Allows remote monitoring teams to mark the injured person in the aerial video stream from a UAV. (b) \humanar: Allows on-scene first response team members to visualize the location of the identified injured victim in (a) using AR glasses or mobile devices. (c) \missionplan: Allows mission planning team to visualize the spatial summary of the mission including the location of UAVs, their aerial video stream, and points of interest identified via (a)}
  \Description{Description}
  \label{fig:teaser}
\end{teaserfigure}
\maketitle

\section{Introduction}
\label{sec:intro}
Technological advancements have enabled Unmanned Aerial vehicles (UAVs) to evolve from manual to semi-autonomous to near fully autonomous operations \cite{floreano2015science}. As a result, they have gained in popularity as surveillance, delivery, photography, and emergency rescue tools, with state of the art emergency response teams exploring the use of smart UAVs for public safety operations such as search and rescue \cite{khan2014information,scherer2015autonomous}, firefighting \cite{jayapandian2019cloud,aydin2019use}, surveillance \cite{basso2018practical,suaruacin2017powering}, and disaster relief \cite{nguyen2019real}.
While technological advances allow UAVs to operate independently, human supervision with timely interventions are still necessary to ensure their ethical and safe operations. Therefore, both humans and smart UAVs are required to work together as a Human-Agent Team (HAT) \cite{klien2004ten} during emergency response \cite{bellamy2017human,cleland2020human,agrawal2020model}.
As an example of this partnership, law enforcement officers in California recently utilized smart UAVs to help them catch a shooter running through the streets \footnote{\url{https://www.nytimes.com/2020/12/05/technology/police-drones.html}}. Officers at their headquarters continuously monitored the live video feed, and relayed details to the police officers on the scene as the smart drone autonomously tracked the shooter through crowded parking lots, strip malls, and narrow streets. While the drone provided a bird's eye view of the evolving scene, effective communication between the off-site officers and the on-site team ultimately played a crucial role in catching the shooter. Similarly, in 2019 French firefighters used two UAVs equipped with thermal cameras to assess the fire scene and distribute fire hoses effectively to fight the fire at Notre-Dame cathedral in Paris \footnote{\url{https://www.theverge.com/2019/4/16/18410723/notre-dame-fire-dji-drones-tracking-stopped-thermal-cameras}}. This massive emergency response operation involved more than 400 people, including firefighters, government officials, police, and municipal workers,  and demonstrated the use of the UAV's onboard cameras, sensors, and computers to augment information collected by on-scene first responders and to provide timely insights to the mission commander and other team members throughout the mission.  It also highlights the need for effective communication amongst humans, smart UAVs, and between humans and UAVs during emergency situations \cite{cleland2020human}. 

To identify communication barriers, Manoj et al. \cite{manoj2007communication} observed the first responders' as they engaged in exercises, drills, and workshops. Apart from the technical challenges associated with providing first responders with a robust communication network during an emergency response, Manoj et al. \cite{manoj2007communication} also identified communication challenges at the social as well as organizational level.  

The social communication challenge is to ensure a coherent interpretation of the situation among emergency responders. Information communicated over the radio suffers from noise in signals, lacks visualization, and is difficult to persist for later analysis. In a separate study \cite{agrawal2020next}, first responders reported that sharing information over the radio in an already noisy environment can lead to an inaccurate perception of a situation. Organizational communication challenges also arise due to information gaps between primary decision-makers such as on-site first responders, remote emergency operations centers (EOC), and local governments when planning emergency responses. Therefore, the social and organizational challenges of communication necessitate improving the way emergency response teams consume and share information \cite{andreassen2020information, bergstrand2009information, guntha2020architectural} during emergency response. This paper aims to address the social and organizational issues of communication during an emergency by supporting rescue team members in geotagging different pieces of information on the scene and visualizing it upon demand. Geo-locatable information can be shared and visualized in different formats such as on 2D maps for mission commanders \cite{domova2020improving}, overlays on video streams for remote team members, and even by displaying virtual objects using augmented reality head-mounted displays for the on-scene first responders\cite{roldan2019multi,roldan2017multi}. UAVs in emergency response also utilize geolocated information to make autonomous decisions during missions. For example, during a river search and rescue mission, a UAV could autonomously deliver a flotation device based on the geolocation of the potential victim \cite{agrawal2020model}.

In this work, we utilize the paradigms of Location-based Augmented Reality to design \system with the intent of helping mission commanders, first responders, and rescue teams to communicate scene information and retain situational awareness as an event unfolds. The design of \system comprises three different interfaces: \Humanar, \Dronear, and \missionplan, all of which empower responders to share and visualize information according to the diverse needs of their specific team roles. \system  automatically computes the geolocation of objects of interest in the aerial video stream captured by UAVs in real-time. These objects can be detected using onboard image recognition models or tagged in the video stream by emergency responders. \system first computes the geolocations of the physical objects in the environment and then places virtual objects at these geolocations in the AR world space. \system also persists the geolocations of objects on a cloud server to make them visually available upon demand across all three UIs. \system represents a pragmatic approach for alleviating the social and organizational challenges of communication during an emergency response.

The main contributions of this paper are: (i) the design and implementation of the \system component: \dronear, as an interface for remote rescue team members to create, share, and visualize AR content during the UAV-driven emergency response mission, (ii) a second interface for On-Scene Response teams: \humanar, which builds upon the design of existing AR interface \cite{campos2019mobile} and adapts them for use in a UAV-driven emergency response system, (iii) feedback from a focus group study that assesses the differences between verbal and visual modes of communicating a scene in a rapidly changing environment, and (iv) recommendations for the design of frameworks for the UAV-driven emergency response system emphasizing visual communication. We conducted multiple studies to learn the feasibility, usability, and end-users' (police officers) perspective of our system. We found that \system offers a pragmatic approach to alleviate the social and organizational challenges of communication during emergency response. \system makes critical geolocation information more accessible, and the visualization of the points of interest (POIs) improves the spatial understanding of the scene among the rescue team members. Further, a team of police officers indicated that \system would complement the existing communication infrastructure (radios) of their emergency rescue teams.

The remainder of the paper is organized as follows. In Section \ref{sec:background}, we provide background knowledge on human-UAV interaction and augmented-reality applications for collaboration. In Section \ref{sec:framework}, we describe \System, including its architecture and implementation details. Section \ref{sec:eval} explains the methodology we used to evaluate \system and present the results from our experiments and a user study. In section \ref{sec:domain_experts}, we describe police officers' perspective on the design of \system. Section \ref{sec:disscusion} summarizes the findings and provide opportunities to explore in future to design AR-Based tools for emergency response. Finally, In Section \ref{sec:limitations}, we discuss the limitations of our work and conclude our discussion in Section \ref{sec:conclusion}.

\section{Background and Related Work}
\label{sec:background}
\subsection{UAV Driven Emergency Response}
UAVs stream the bird's eye view of emergency situations and can provide valuable information to the rescue team \cite{jones2016elevating}. Khan et al.,  \cite{khan2019exploratory} interviewed firefighters and learned that communicating effectively with remote emergency rescue teams is critical in the design of a UAV driven emergency response system. Similarly, Lopez et al., \cite{lopez2017dronealert}  demonstrated that using live video footage of the incident from UAVs enables firefighters to devise strategies to mitigate damage ahead of time. Similarly, our work leverages real-time aerial video streams to distribute scene information. In particular, we focus on designing interfaces for demand-based visualization of the scene information to improve the situational understanding of various rescue teams.

Agrawal et al.,  \cite{cleland2020requirements,agrawal2020model,agrawal2020next} conducted multiple participatory design sessions with firefighters to explore the role of smart UAVs' assistance in emergency situations. They also provided guidance for designing interfaces for UAV-based emergency response systems, and emphasized the need to support multi-user collaboration between humans and smart UAVs \cite{cleland2020human, abraham2021adaptive}. While this prior work has  studied the importance of humans and multiple UAVs  collaborating under various circumstances, the new work described in this paper presents a novel interaction technique based on annotating aerial video stream with POI information and visualizing it across role-specific interfaces, to improve human-UAV collaboration during emergency response scenarios.

\subsection{Augmented Reality for Collaboration}
Augmented Reality (AR) technology augments the humans' perception of the real world by adding virtual elements to it. Location-based AR experiences \cite{paucher2010location} rely on the location of the device to determine where and how to place a virtual object within the real world, whereas vision-based AR relies on known image markers \cite{kato1999marker} or complex scene analysis algorithms to augment the real world with virtual objects \cite{gauglitz2014world,carozza2014markerless}. In contrast, Remote Augmented Reality (RAR) extends location or vision-based AR experiences by augmenting videos received from remote sources with 3D models and graphics in real-time.  The usability of RAR  has been explored widely for remote assistance applications in the manufacturing industry for connecting end-users to remote experts for product maintenance \cite{masoni2017supporting} \cite{mourtzis2017augmented} \cite{mourtzis2017cloud} \cite{schneider2017augmented}, and in the healthcare industry for remote surgical guidance, and teleportation \cite{wang2017augmented} \cite{mather2017helping} \cite{rojas2020system}. Similarly, Unal et al., \cite{unal2020distant} demonstrated the use of RAR for delivering a lively experience of cultural heritage to tourists. In contrast, we utilize RAR to share and communicate scene information to team members.

In the context of emergency response, researchers discovered AR's suitability for communicating information within and among rescue teams during earthquakes \cite{leebmann2004augmented}, fire outbreaks \cite{wani2013augmented}, and medical emergency services \cite{chi_2021_hwd}. The THEMIS-AR \cite{nunes2018augmented} is an AR mobile application designed to improve emergency responders' scene perception by overlaying context-relevant information, such as distance, time, and position of POIs on their scene perception via a head-mounted display. 
Further, Campos et al., \cite{campos2019mobile} extended THEMIS-AR to study AR user interfaces and  provide guidelines for designing AR applications for emergency response. However, the scene information in THEMIS-AR is geo-referenced manually, and the AR application overlays that same information on the first-person views of onsite first-responders, such as the number of casualties, distance to the nearest hospital, or fire location. In contrast, this paper builds upon previous work for UAV based emergency response systems by 1) automatically computing the Geo-location of POIs from their image pixel coordinates in the aerial video frame, 2) allowing remote rescue teams to add, track and visualize POI data in real-time as overlays over videos streamed by UAVs via \Dronear, and 3) allowing on-site first responders to  visualize dynamically identified scene information by other rescue teams or UAVs in the first-person view via \Humanar. Further, \system also allows decision-makers to continuously visualize the points of interests on 2-D maps via its \missionplan interface as the situation evolves.

\section{Collaboration Framework: Rescue-AR}
\label{sec:framework}

\begin{wrapfigure}{r}{0.5\textwidth}
    \centering
    \includegraphics[width=0.5\textwidth]{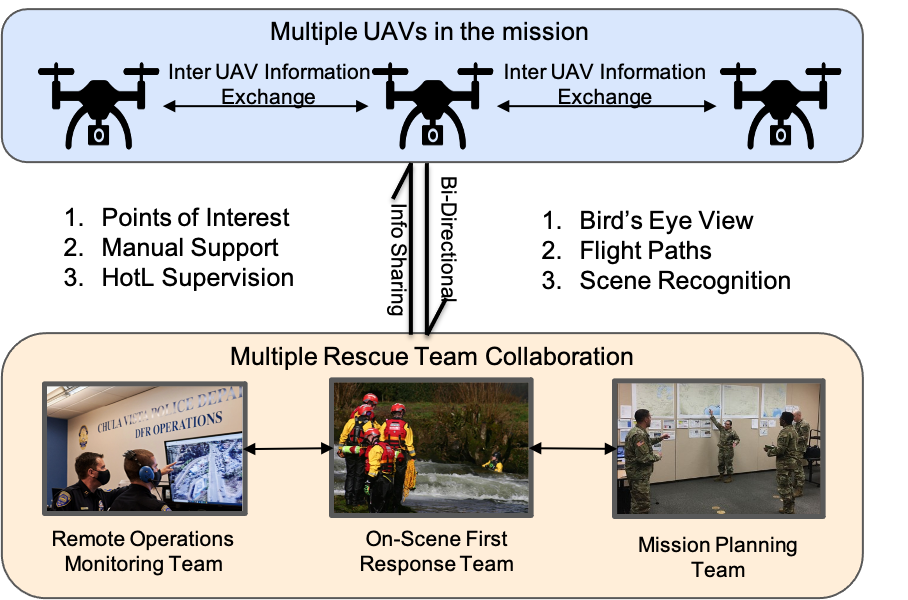}
    
    \caption{Bi-Directional Information Exchange among Multiple Humans and Multiple UAVs During Emergency Response}
    \label{fig:Framework_Overview}
\end{wrapfigure}

The purpose of \system is to enable diverse human rescue teams and UAVs deployed on a mission to share and visualize information. \system provides support across rescue teams in three primary roles: first,  {\it On-Scene First Responders} responsible for assessing the scene closely, and performing ground activities such as rescuing victims or arresting suspects; second, {\it Mission Planners}, whose primary responsibility is to develop strategies for rescue missions and impart plans of actions to individual teams; and third, the {\it Remote Monitors}, whose primary responsibility is to closely monitor the rescue operation and provide informational support to on-scene responders and mission planners. Smart UAVs deployed in the mission use their on-board vision and computation capabilities to independently carry out tasks,  such as surveying impacted areas, tracking people, or detecting and locating objects of interest to rescue teams. Information collected by the UAVs can be used by human responders to gain increased situational awareness throughout the operation. At the same time, human operators supervise the UAVs and, when needed, provide them with information garnered from the ground to help them enact their tasks. As an example, a human might provide the GPS coordinates of a victim trapped in trees at the edge of the river so that the UAV can deliver a flotation device. Figure \ref{fig:Framework_Overview} shows the kinds of collaboration and bidirectional exchange of information that is possible between human team members and UAVs during emergency response.  

\system provides three interfaces that  rescue teams can use to share and visualize information. We developed these interfaces using several iterative prototype design, development, and evaluation cycles, to allow for continual refinement.  To help achieve consistency of information shared across all three interfaces, we used Marinova's map symbol system \cite{marinova2018new} to represent points of interest such as a disaster, infrastructure, people, and operational sites.

We now describe the design and features of each interface: \Missionplan, \Dronear, and \Humanar \footnote{Demo Videos: \url{https://www.youtube.com/playlist?list=PLU5_FMa4WPVgTSWHJmmdHVK-Eu3HvZLRS}}. 
\missionplan is a classic web-based application that builds upon DroneResponse \cite{agrawal2020next} to visualize the POIs on 2D maps. The \dronear and \humanar  are standalone Location-Based Augmented Reality applications developed in the Unity \cite{Unity:online} environment for iOS devices using ARKit \cite{AR-kit:online} platform. We study the usability of our interfaces for the collaboration across teams in the domain of multi-UAV systems. 

\subsection{\dronear}
\label{interface-dronear}

\dronear leverages Location-Based AR and relies upon geolocations to augment the video stream with POIs. Remote monitors use \dronear to monitor the aerial video feeds of UAVs involved in the mission and to annotate them with identified POIs throughout the mission. Once created, these POIs appear as overlays on the aerial videos of any UAV flying in the POI's vicinity.  In addition to human-created annotations, \dronear can also leverage onboard vision capabilities (e.g., YOLO \cite{yolo_redmon2016you}) to automatically detect and annotate additional objects of interest. However, while YOLO and other similar vision models are able to recognize diverse objects, such as people, vehicles, and animals, recognizing a broad set of unknown POIs presents an Unknown-Unknown problem, and is the reason that we designed \dronear to also support manual tagging by operators. Regardless of how the annotation is created, the pixel coordinates of the video frame representing the POI need to be transformed into a geolocation to augment its corresponding virtual object in the streamed aerial video. In this paper, we focus on the human aspects of annotating the aerial video stream. 

\subsubsection{POI Augmentation}
\dronear simultaneously collects the following four pieces of information when a POI is marked in a video frame: 1) the GPS position of the UAV, 2) orientation of the UAV (Roll, Pitch, Yaw), 3) orientation of the on-board camera (Roll, Pitch, Yaw), and 4) the video frame itself. This data is used to transform each pixel coordinate in the frame into geolocations. 
The AR World tracking camera of \dronear leverages this data to mimic the position and orientation of the UAV and camera onboard the drone in the virtual environment. Further, we also configure the Field-of-View of the AR camera to be the same as that of our UAVs to maintain consistency between physical UAV attributes and the AR application. In order to determine the final perspective of the onboard UAV camera inside the AR environment, \dronear employs two transformations: UAV Position and Orientation, and Gimbal  Orientation. Both of these transformations are applied in the same order at each update in the aerial video frame. \dronear locates the annotated region of the frame in order to compute the geolocation of the POI, and then uses this computed geolocation along with AR world tracking to render and track the virtual object on the video stream. Finally, it computes the geolocations of the POIs from their pixel coordinates in the aerial video frame.

\subsubsection{Geo-Location of POIs}

GPS uses the ECEF (Earth-Centered, Earth-Fixed) Cartesian coordinate system as its primary coordinate system and drives all other coordinates from it. Therefore, we perform all our calculations in the ECEF system. The computation is described in the following steps:
\begin{itemize}
    \item \textit{Transform UAVs' GPS to ECEF}: First, we transform the incoming GPS location from the UAV into its corresponding ECEF coordinates using a standard computation suggested by Zhou, \cite{zhou1999sensor}. We define the ECEF coordinate of the UAV's location as $ECEF_{UAV}$ containing $X_{UAV}$, $Y_{UAV}$, and $Z_{UAV}$ as three dimensions of the ECEF coordinate system. 
    
    \item \textit{Compute ECEF of the POI}: Second, we leverage the \emph{ray tracing} feature of the ARKit platform, which returns a ray going from the AR camera through a screen point (representing the POI) to compute the position of the POI in the AR world space. This position represents the ENU coordinate of the POI with respect to the position of the UAV. We therefore use the following equation to compute the ECEF coordinates of the POI:

\begin{equation}	
	\left[\begin{array}{c}
		ECEF_{POI}
	\end{array}\right]=
	\left[\begin{array}{c}
		X_{POI} \\
		Y_{POI} \\
		Z_{POI}
	\end{array}\right]=\left[\begin{array}{ccc}
		-\sin \lambda & -\sin \phi \cos \lambda & \cos \phi \cos \lambda \\
		\cos \lambda & -\sin \phi \sin \lambda & \cos \phi \sin \lambda \\
		0 & \cos \phi & \sin \phi
	\end{array}\right]\left[\begin{array}{l}
		x \\
		y \\
		z
	\end{array}\right]+\left[\begin{array}{c}
		X_{UAV} \\
		Y_{UAV} \\
		Z_{UAV}
	\end{array}\right]   
\end{equation}

Here, $\phi$ and $\lambda$ values are the geographic raw latitude and longitude of the UAV location, while  x, y, z represent the position of the POI in the AR world space.

\item \textit{Transform ECEF of the POI to GPS}: Finally, we transform the $ECEF_{POI}$ into its corresponding Geo-Location using the standard algorithm provided by \cite{zhu1994conversion}, Zhu. 
\end{itemize}

After computing the geolocation of the target object, \dronear pushes the geolocation to a backend component for persistence and for distribution to other UAVs and rescue team members participating in the mission.

\begin{figure}[htbp]
  \centering
  \subfloat[The user specifies the POI (Injured Person) by marking the area of the screen with an appropriate symbol on the aerial video stream of UAV-1 via \dronear. \system computes the geolocation of the point of interest and uses it to augment the aerial video stream with the symbol. ]{\includegraphics[width=0.45\columnwidth]{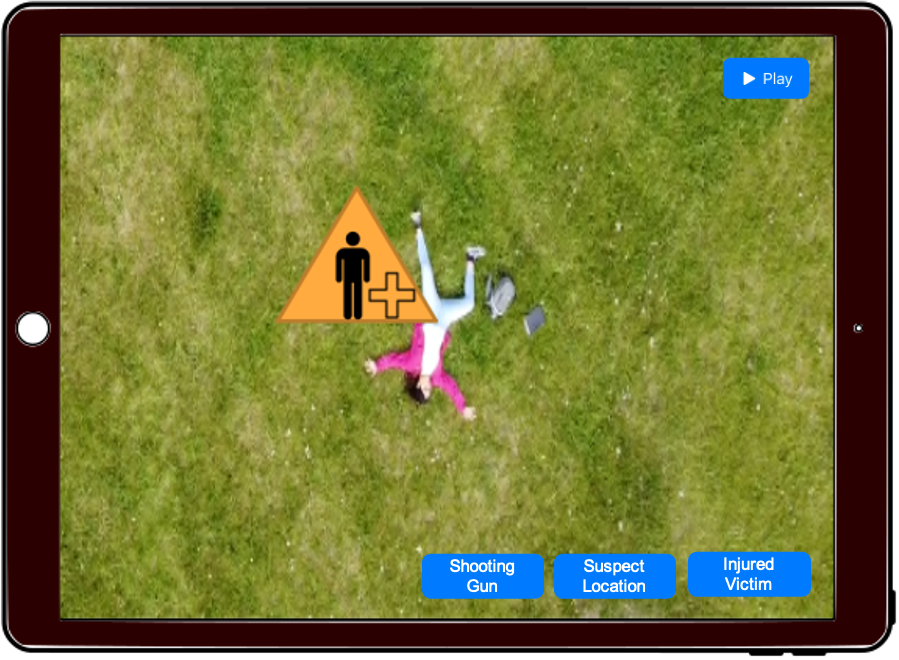}
  \label{fig:remote-team-interface-drone-1}}
  \hfill
  \subfloat[The system persists the geolocation of the POI (Injured Person) identified in the aerial video stream of the UAV-1. Therefore, \dronear uses it to augment the video stream of UAV-2 in the same mission with an appropriate marker when its camera looks towards the POI's geolocation.  ]{\includegraphics[width=0.45\columnwidth]{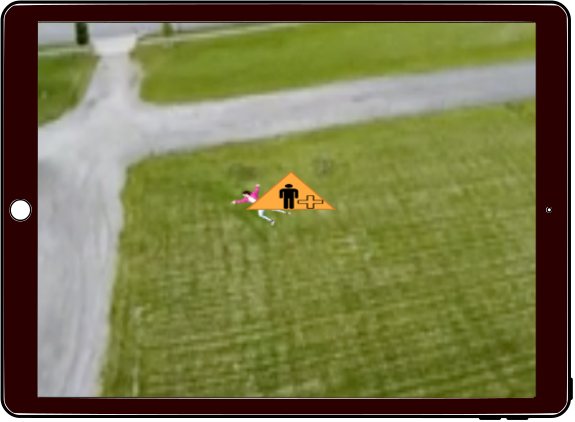}
  \label{fig:remote-team-interface-drone-2}}
  \caption{POI annotation and visualization over aerial video streams of multiple UAVs via \dronear interface}
\end{figure}

\subsubsection{Multi-UAV POI Visualization}
To preserve the geolocation information in a way that can be shared with other UAVs participating in the mission, all symbols corresponding to POIs are persisted in the AR world space of \dronear based on their geolocation. As all UAVs share the same AR world space of \dronear and as each UAV has its own AR camera, we are able to create a shared augmented reality \cite{peitso2020promise} environment. To support the POI visualization for multi-UAV emergency response missions, we leverage the AR world tracking feature of the ARKit platform to render the POI annotated in the aerial video stream of one UAV into the aerial video stream of other UAVs participating in the same mission. Figures \ref{fig:remote-team-interface-drone-1} and  \ref{fig:remote-team-interface-drone-2} illustrate the multi-UAV POI visualization whilst searching for an injured person. The sequence of events for multi-UAV POI visualization is as follows; first, the victim is detected in the aerial video stream of a UAV (e.g., UAV-1, Figure \ref{fig:remote-team-interface-drone-1}) and a user marks the victim on the video stream; second, \system computes the geolocation of the injured person and stores it in the shared AR space. Finally, the AR cameras of other UAVs (e.g., UAV-2, Figure \ref{fig:remote-team-interface-drone-2}) track the same AR space and render the POI overlay into their own video streams when the POI appears within their camera's field of view.

\subsection{\humanar}
\label{interface-humanar}

\humanar is a Location-Based AR application that allows the On-Scene First Response team to visualize geolocated scene information from the first-person perspective using head-mounted displays, mobile devices, or AR glasses. \humanar creates its own AR space; however, it constantly synchronizes the geolocations of POIs from the AR space of \dronear using the backend component of the \system over secured Wi-Fi channels. The AR camera of the \humanar application mimics the GPS position, orientation, and camera properties such as Field-of-View of the first responders' device in the AR world space. This enables \humanar to track the AR world space according to the movements of the first responder in the real world. \humanar also  leverages ARKit's \cite{AR-kit:online}  world tracking feature to augment the POI symbol in the first responders' view based on geolocations of the POI and the first responders' device. The \humanar interface extends Campos et al., THEMIS-AR \cite{campos2019mobile} design for use in a multi-UAV domain. While THEMIS-AR requires humans to manually provide the geolocation of the POI in the form of GPS coordinates, our solution lifts this limitation by automatically computing the geolocations of POIs based on annotations in the video stream. Additionally, \humanar leverages web-socket connections over Wi-Fi and LTE channels to exchange data such as geolocation of POIs with other emergency response tools. Hence, the extensible design of \humanar allows it to receive GPS locations of POI at run-time from UAVs, \missionplan, \dronear, and can work in conjunction with other emergency rescue tools. Figure \ref{fig:firstresponse-team-interface} shows the \humanar application interface.

\subsection{\missionplan}
\label{interface-missionplan}

\begin{figure}[!tbp]
  \centering
  \subfloat[\missionplan UI providing access to location, flying pattern, status of four UAVs searching an area. The UI also allow visualizing the POIs (Injured Victims, Guns) on 2D maps.]{\includegraphics[width=0.45\columnwidth]{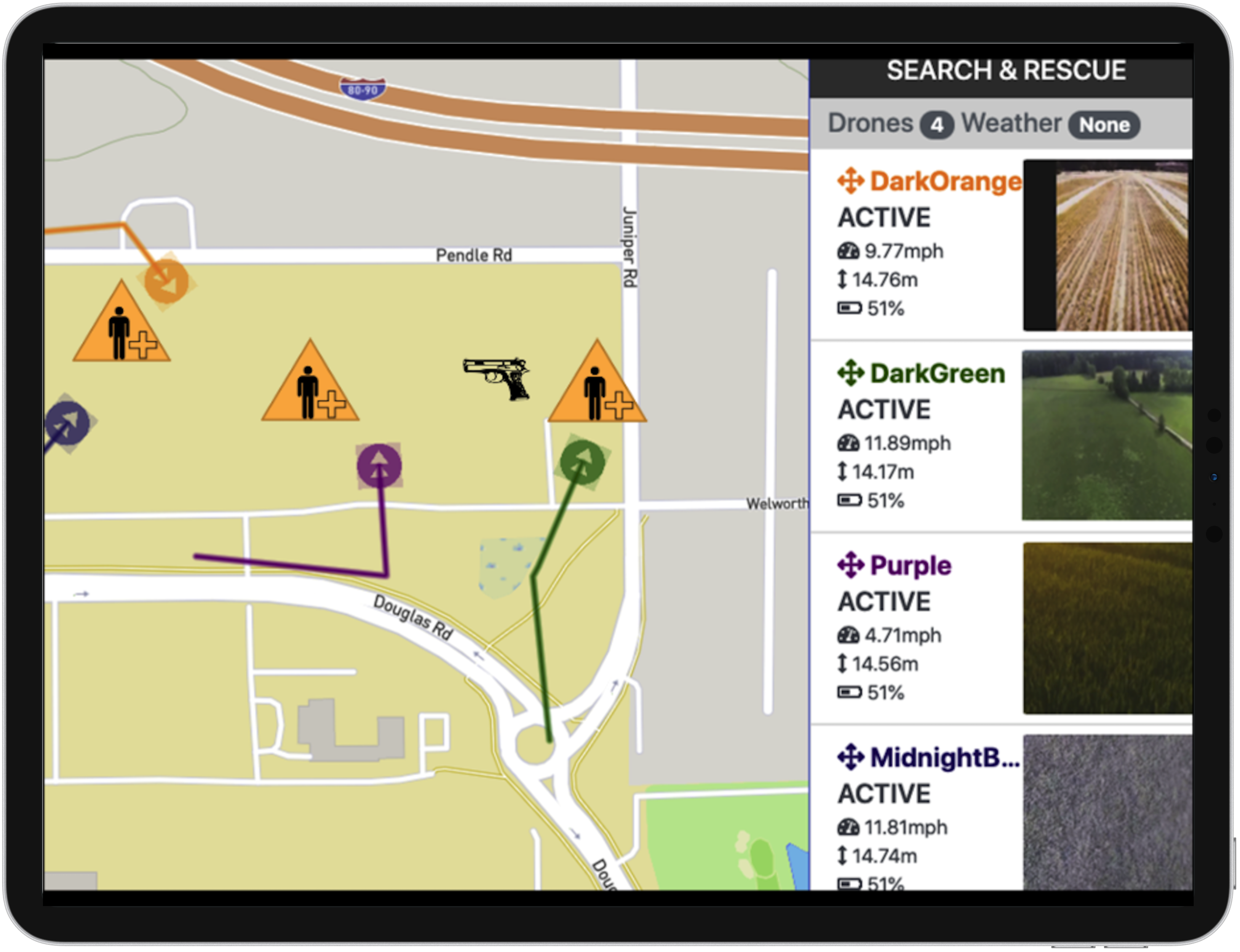}
  \label{fig:mission_planning_team_interface}
  }
\hfill
  \subfloat[\humanar interface allowing on scene rescue team members to visualize geolocated markers using AR Glasses and AR mobile apps]{\includegraphics[width=0.45\columnwidth]{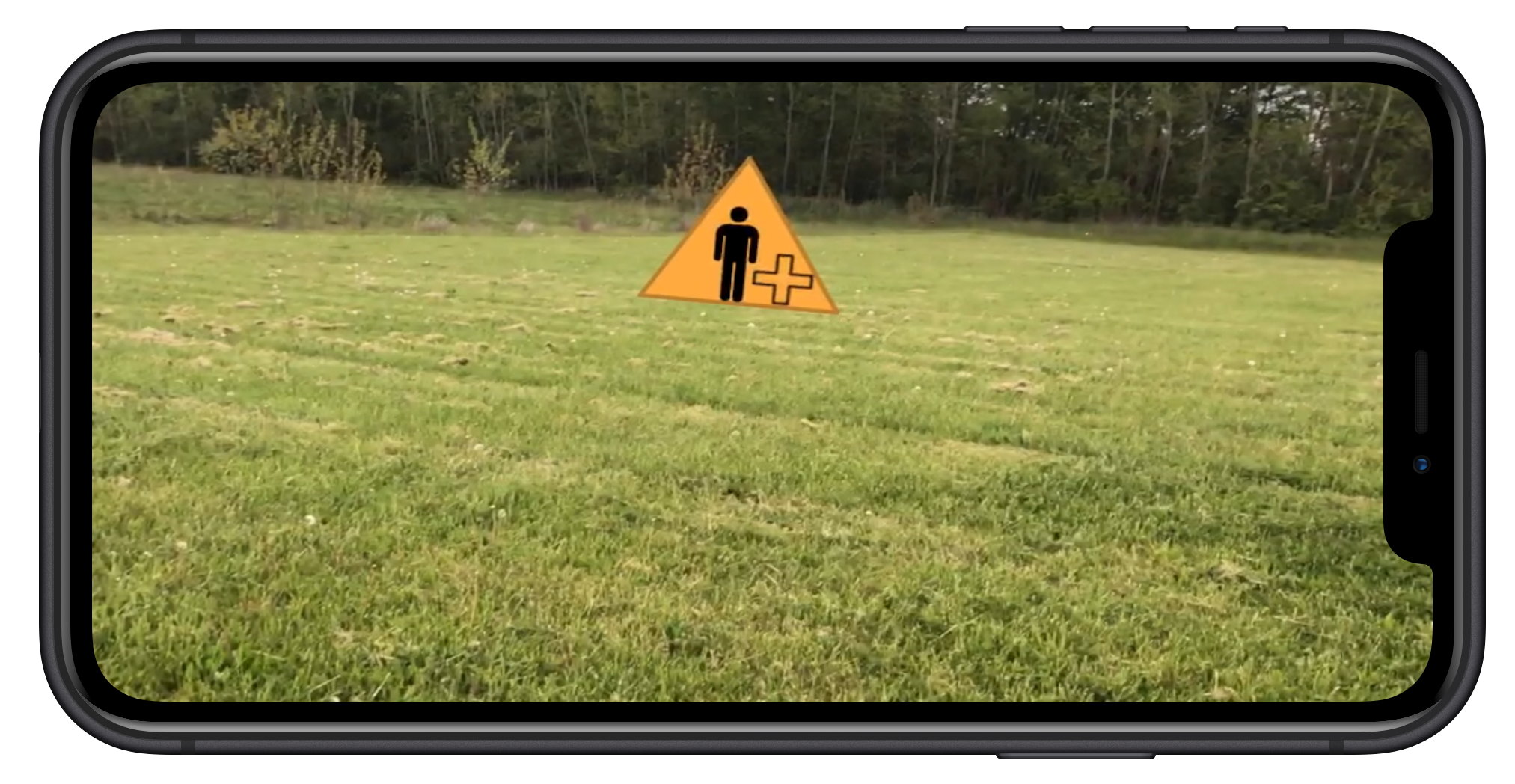}
  \label{fig:firstresponse-team-interface}}

  \caption{The POIs identified using the UAVs' aerial video stream are shared across the rescue team members for visualization via \dronear and \missionplan}
\end{figure}

The Mission Planning Team leverages the \missionplan interface to visualize the overview of the mission, allowing the team to plan and discuss different execution strategies as the mission unfolds. The \missionplan interface is implemented as a Web application using the Angular \cite{Angular:online} web framework and displays video streams and UAV location and status, while \missionplan is deployed on a web server, where it enables each member of the Mission Planning team to access the status of the mission on their devices. It extends the design of previous multi-UAV interfaces from the literature \cite{agrawal2020next} by (i) receiving information from multiple rescue teams and visualizing it on 2D maps, and (ii) broadcasting Geo-Locations of POIs to UAVs and other stakeholders of the mission. Figure \ref{fig:mission_planning_team_interface} shows the \missionplan interface. The team interacts with the user interface to perform tasks such as plan flight routes, assigning tasks to specific UAVs, and annotating maps with scene information using labels and icons. Whenever a scene update is received for a new piece of geo-locatable information, \missionplan synchronizes the information to all devices and immediately displays it on the map, reducing the need for redundant communication amongst distributed members of the Mission Planning Team. 

\subsection{Collaboration}

\system supports collaboration through a back-end component that persists the geolocation of all POIs annotated by either the rescue team members or UAVs themselves. The back-end is built using Node.js, and deployed as a web service on a private cloud server using web sockets to broadcast data in real-time. The back-end component tracks the geolocation, type of POI, information source (e.g., rescue team ID), and other properties of the shared data. Each member of the rescue team is equipped with a mobile device which communicates with the node.js web service to synchronize the latest information in the UI. 
We describe two primary scenarios for illustrating the usability of \system in visualizing the situational information and supporting collaborative work whilst using UAVs for emergency response.

\subsubsection{Human Knowledge Sharing}
In the river search and rescue scenario, autonomous UAVs search the river for drowning victim(s) whilst continually streaming video. Team members, including the Remote Operations Monitoring Team, examine aerial video streams for information that could be important for the search mission, such as clothing, a capsized canoe, or submerged branches in which a victim might get trapped. 
While the UAV video streams can provide mission-critical information; precisely communicating this information across multiple rescue teams remains challenging due to the difficulty of verbally describing geographic locations. 
As highlighted in Figure 2, the \dronear interface, supported by the AR environment, is designed to enable high degrees of collaboration between team members monitoring the video stream and those enacting actual operations at the scene of the incident.  The enhanced situational awareness provided by \dronear enables timely planning and more effective mission execution.  

\begin{figure}[!tbp]
  \centering
  \subfloat[\textbf{Human Knowledge Sharing}: Remote Operations Monitoring Team identifies an injured person via Video Streams. The collaboration framework computes the geolocation of the injured person and shares it with On-Scene First Response Team and Mission Planning Team]{\includegraphics[scale=0.42]{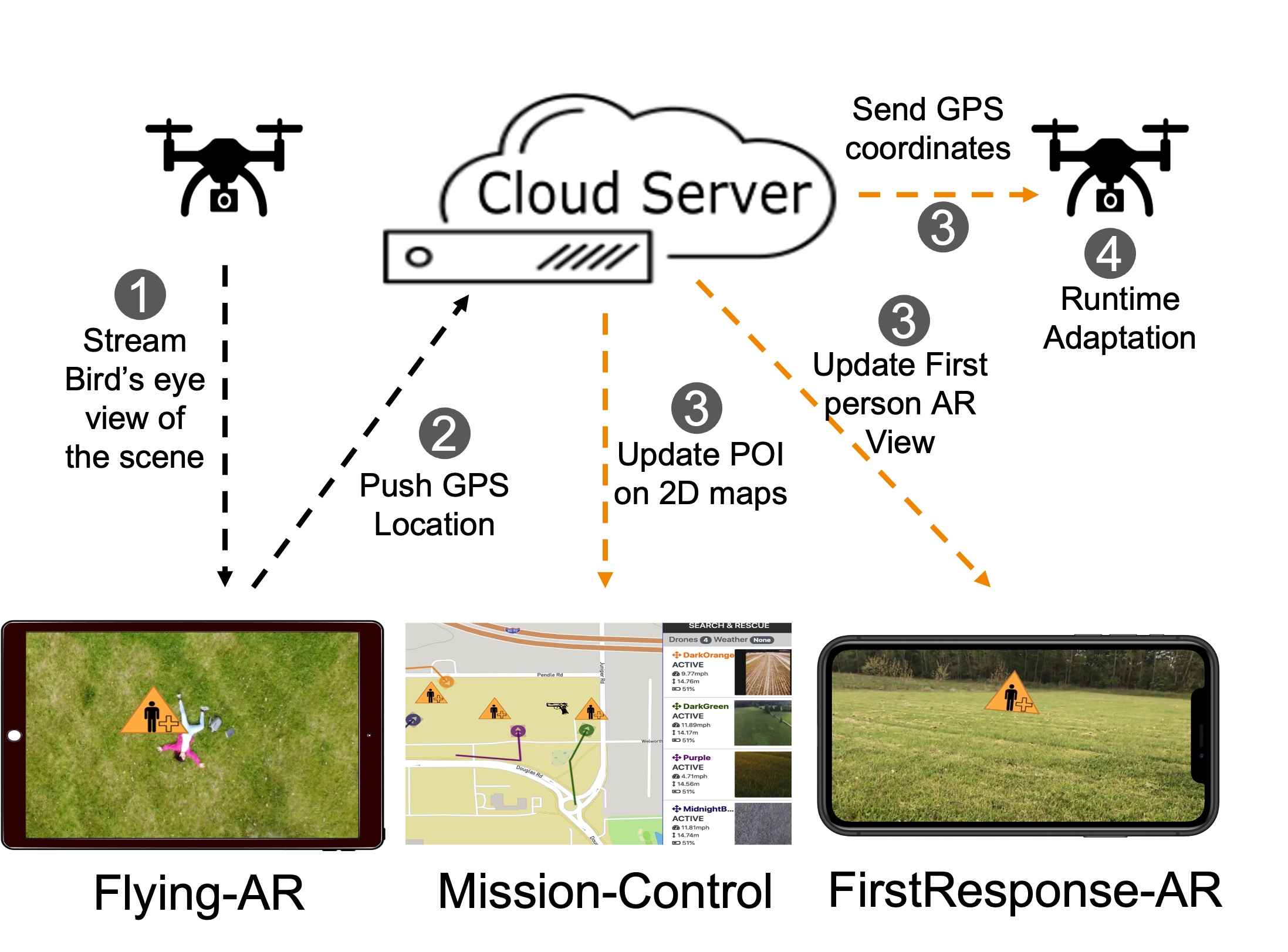}
  \label{fig:s1}}
  \hfill
  \subfloat[\textbf{UAV Knowledge Distribution}: The vision model onboard a UAV first detects a potential victim in the scene. \system computes the geolocation of the person in the frame and pushes it for visualization on 2D maps, \Dronear, and \Humanar ]{\includegraphics[scale=0.42]{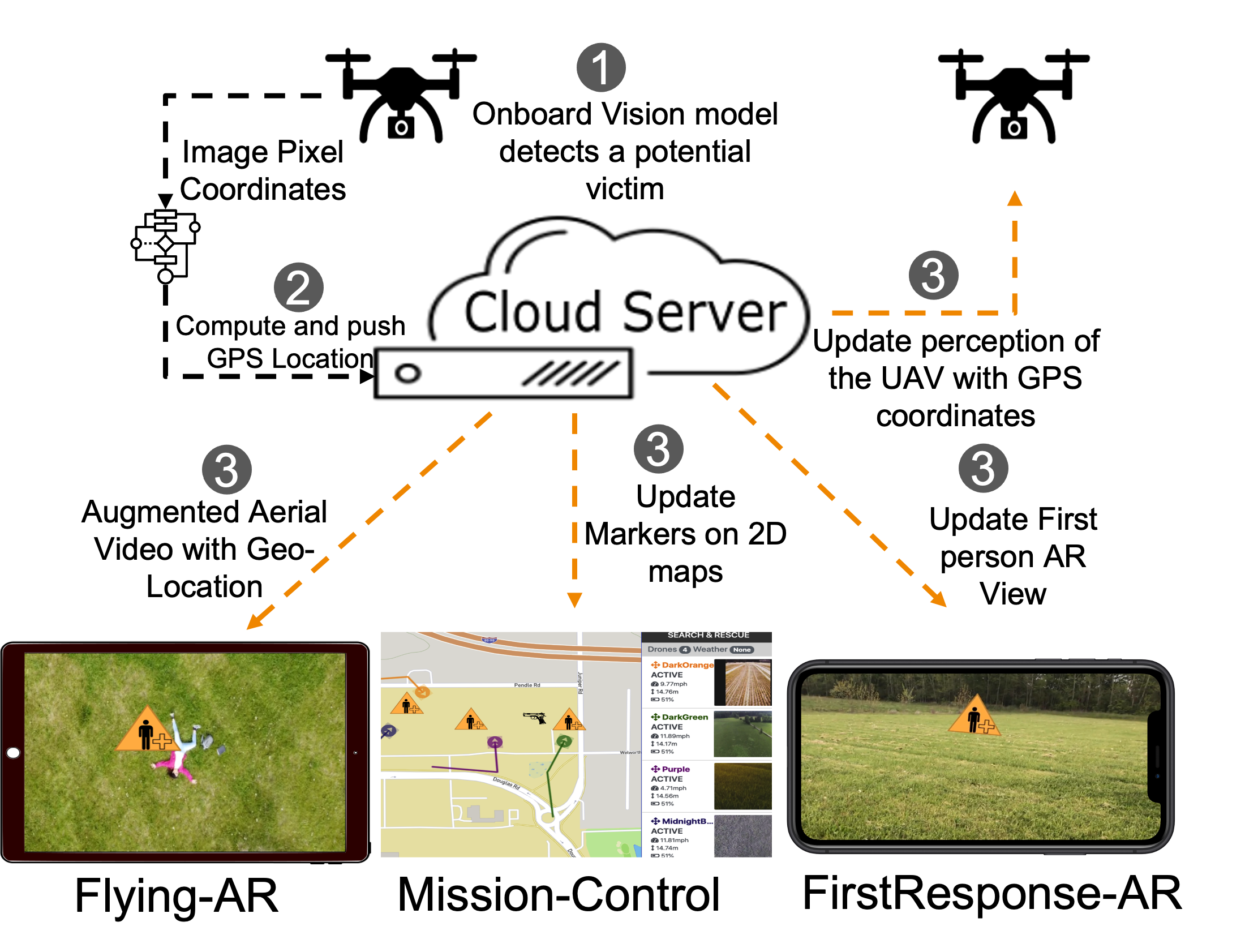}
  \label{fig:s2}}
  \caption{Two primary use cases of \System during UAV-driven Emergency response}
\end{figure}

The geolocation of all POIs is shared with all UAVs participating in the mission through the same back-end component of \System. Given a certain degree of autonomy, the UAVs can leverage this information to adjust their flight plans and actions accordingly. In summary, \system allows key mission information to be shared in close to real-time to all mission participants, including UAVs, in a format that facilitates comprehension and actions. Figure \ref{fig:s1} provides an overview of the scenario.

\subsubsection{UAV Knowledge Distribution}
Companion computers onboard the UAVs enable them to act and perform tasks independently. These companion computers, such as Nvidia Jetson Nano \cite{JetsonNa82:online}, run complex planning \cite{allaire2008fpga,zhou2020trajectory}, vision \cite{timotheatos2018vision,yang2019imgsensingnet}, and adaptation algorithms \cite{zhang2018novel} that allow UAVs to sense, plan, and act in the environment. During a river search and rescue operation, the onboard companion computer performs the computations described in section \ref{interface-dronear} to transform the pixel coordinates of the point of interest recognized by the onboard vision algorithms into its corresponding Geo-location. The UAV pushes geolocation of POIs to the back-end component for broadcasting. This allows a UAV to directly share information with all stakeholders, including other UAVs participating in the mission. The mission planning team visualizes the POIs in the form of icons on 2D maps. The world tracking feature of the Flying AR interface enables members of the remote operations monitoring team to visualize the UAV recognized POIs on the aerial video streams of all UAVs. Likewise, the on Scene First Response team members visualize the incoming information from UAVs in the first-person view on the ground. Figure \ref{fig:s2} provides an overview of the scenario.

\section{Experiments and Analysis}
\label{sec:eval}
The objective of our study was two-fold: (i) to evaluate the effectiveness of the \system in computing the geolocations of POIs in the aerial video stream, and (ii) to analyze the usability of \dronear in creating and sharing the AR content, and then consuming the AR content as overlays on aerial video stream to understand the scene, and communicating the scene information to rescue team stakeholders. We conducted our user studies after approval from the local institutional research board. For aerial video analysis, we leveraged a commercial grade UAV equipped with a 3-axis Gimbal and an RGB camera to record aerial videos of our testing site. Table \ref{tab:uav-spec} describes the relevant specifications of the UAV used in all of our studies.

\begin{table}[]
    \centering
    \begin{tabular}{|c|c|}
\hline
\multicolumn{2}{|c|}{\textbf{Aircraft}} \\
\hline
Manufacturer & DJI \\

\hline
Model & Mavic Mini 2 \\
\hline
Takeoff Weight & 249 g \\
\hline
Dimensions & 245 X 289 X 55 mm (LXWXH) \\
\hline

Max Flight Time & 30 minutes \\
\hline
Hovering accuracy (GPS based) & ±0.5 meters \\
\hline
\multicolumn{2}{|c|}{\textbf{Gimbal}} \\
\hline
Stabilization & 3-axis \\
\hline
Pitch / Tilt range & -90 to 0 $^{\circ}$ \\
\hline
\multicolumn{2}{|c|}{\textbf{Camera}} \\
\hline
Sensor Size & 1/2.3" CMOS \\
\hline
Effective Pixels & 12 MP \\
\hline
Lens FOV & 83 $^{\circ}$ \\
\hline
\end{tabular}
    \caption{The detailed specifications of the UAV used in our studies.}
    \label{tab:uav-spec}
\end{table}

\subsection{Geo-Location Computation Analysis}
\label{sec:accuracy_analysis}
We conducted a preliminary analysis to identify the suitability of \system in real world conditions. Specifically, we assessed its ability to compute geolocations of objects based on the image pixel coordinates in their aerial video frames.
\subsubsection{Data Collection}

\begin{figure}
    \centering
    \includegraphics[width=\columnwidth]{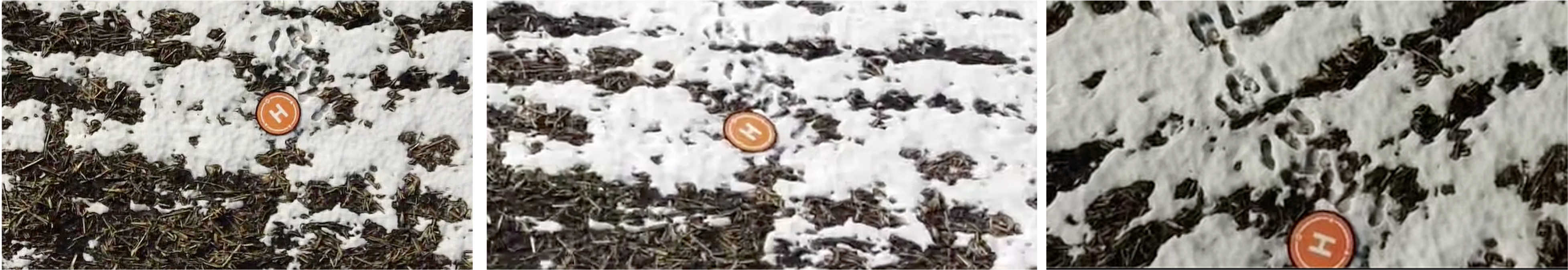}
    \caption{Data Collection: (Left) UAV flying on top of the target object at high altitude (Middle) UAV flying at a distance from the target object, but its tilted camera angles capture the target object in the center of the screen (Right) UAV flying closer to the target object captures it at the bottom of the screen}
    \label{fig:acc-analysis}
\end{figure}

For preliminary testing purposes, we manually operated the UAV so that we could capture diverse views of a target object in the aerial video frame, varying the drone's altitude, position, orientation (yaw, pitch, roll), and gimbal orientation. Figure \ref{fig:acc-analysis} presents a few samples from our video collection showing the target object from various drone altitudes, camera angles, and screen positions. In total, we collected three aerial videos of the scene along with their flight logs, with an average flight time of 3 minutes 4 seconds. Finally, to create approximate `ground truth' about the actual location of the target object, we used \textit{Coordinates}, an iOS mobile application \cite{Coordin89:online} to collect its GPS coordinates and refer to these as {\it measured} coordinates in the remainder of this discussion. For analysis purposes, we compared the measured geolocation of the target object with its computed geolocation.

\subsubsection{Analysis of Geo-Tagging Accuracy}

\begin{figure}[htbp]
  \centering
  \subfloat[Average Horizontal accuracy of \system in computing the geolocation of the target objects in the aerial video stream]{\includegraphics[width=0.5\columnwidth]{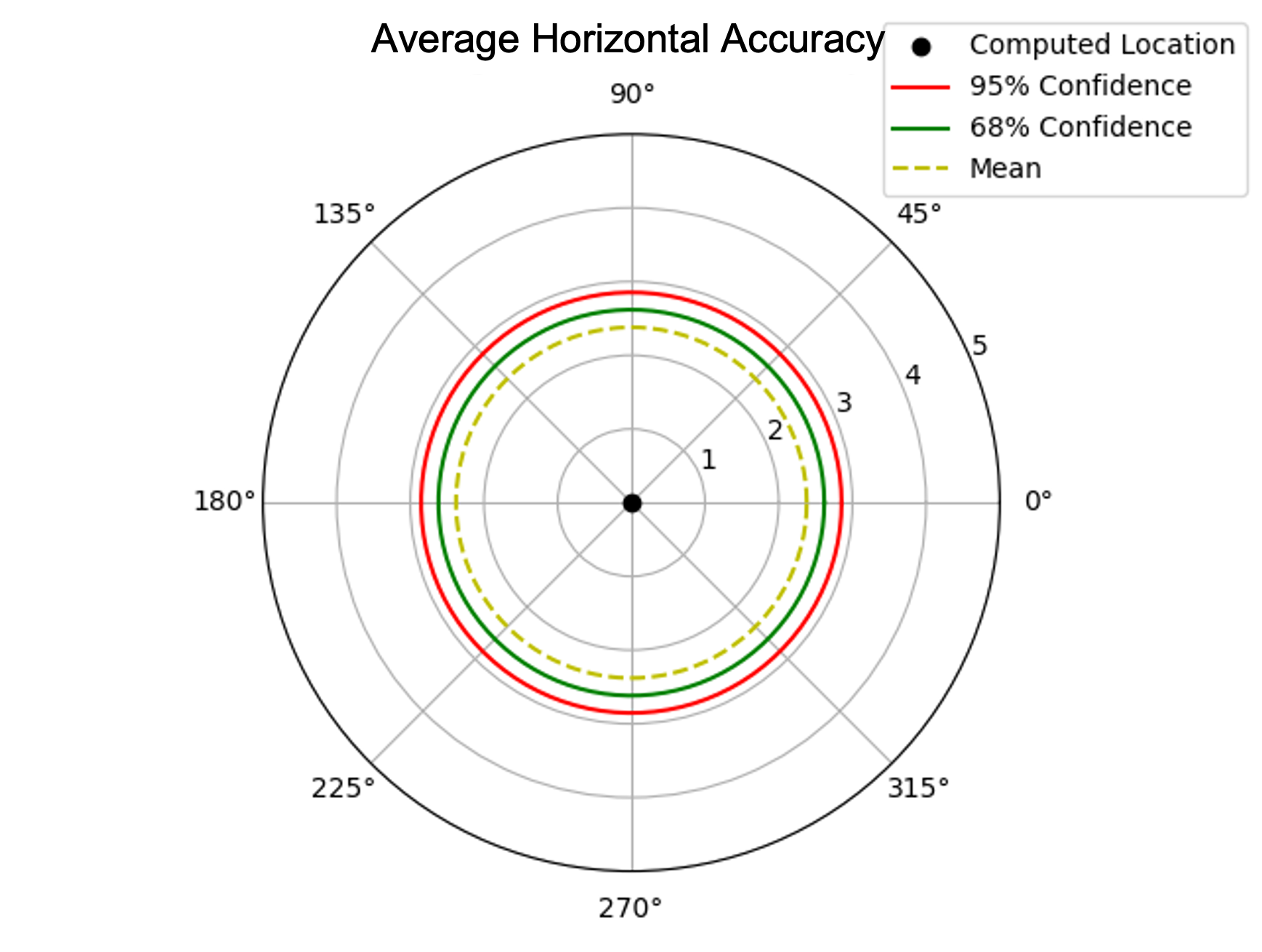}
  \label{fig:avg_horizontal_accuracy}}
  \hfill
  \subfloat[Set-up of the synthetic shooting event to record the aerial video of the scene using our UAV for testing purposes as well as demonstration purposes]{\includegraphics[width=0.35\columnwidth]{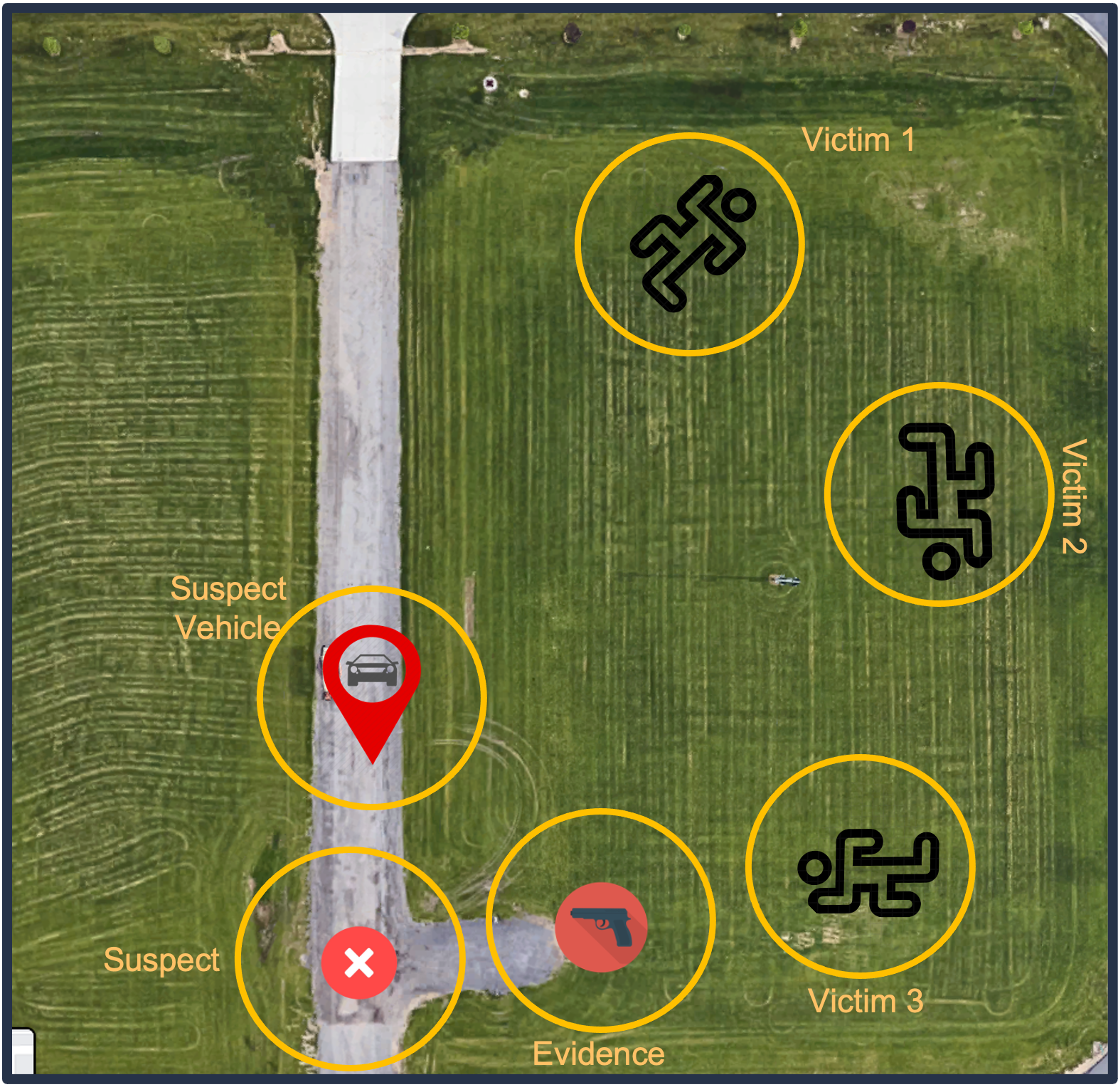}

   \label{fig:crime_scene}}
\caption{\textbf{Left}: Average Horizontal Accuracy; \textbf{Right}: Planned Crime Scene for conducting usability analysis}
\end{figure}

We started by annotating the target objects in the aerial video stream using \dronear. The first author created 91 annotations of target objects in the aerial videos, and recorded their computed geolocations for analysis purposes. We used distance between the calculated and measured co-ordinates of the target object to estimate the horizontal accuracy of our computed geolocations. Figure \ref{fig:avg_horizontal_accuracy} shows the mean horizontal accuracy of the calculated geolocations of the target objects. Each circle in  Figure \ref{fig:avg_horizontal_accuracy} indicates the estimated distance (in meters) at which the measured geolocation of the target would lie from the computed geolocation (center). The red circle indicates that 99\% of the time, the distance between the computed geolocation and the actual geolocation will be less than 3 meters, whilst the green circle indicates that 68\% of the time, the computed geolocation will lie within 2.6 meters of the measured geolocation. 

Since our POI geolocation computations depend on the GPS location of the UAV as described in Section \ref{interface-dronear}, we analyzed the efficiency of our approach under varying UAVs' GPS signal strength. The satellite count available for the GPS device onboard the UAV determines the quality of GPS signal reception. According to the flight logs analysis tool called \textit{AirData} \cite{DroneDat53:online}, availability of at least 13 satellite signals is required for safe flying, and a satellite count below 13 is considered to create poor or dangerous situations for flying. Therefore, we separated our analysis based on UAV's lowest satellite reception count; 15 (excellent flying conditions) and 13 (good flying conditions). 

\begin{figure}
    \centering
    \includegraphics[width=\columnwidth]{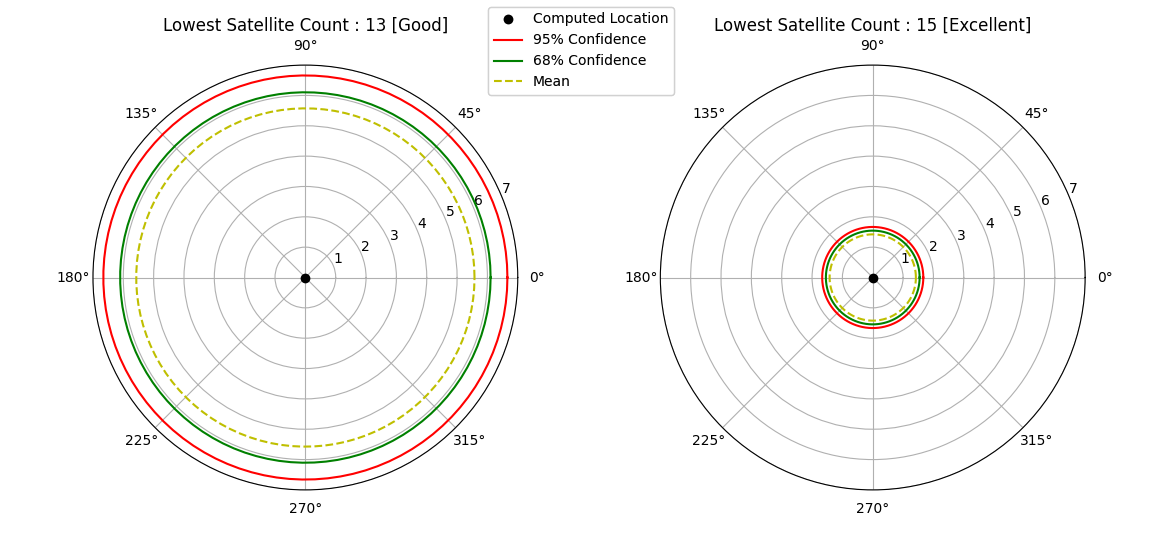}
    \caption{Comparing the accuracy of POI geolocation computation under varying GPS signal quality of UAVs. UAV receives a minimum of 13 satellite count (Left) and 15 satellite count (Right) to estimate its GPS position in the air.}
    \label{fig:accuracy_by_satellites}
\end{figure}

First, we found that the accuracy of the target geolocation computation improves as the number of available satellite count increases. Second, we observed that under excellent signal strength (i.e., satellite count above 15), the mean accuracy is within 2 meters. However, the mean accuracy degrades  to within 7 meters when the satellite counts drop to 13. As a result, an object marked in the video stream of one UAV can be expected to appear within 2 meters of another UAV's video stream, provided there are at least 15 satellites available for UAVs' GPS.  Figure \ref{fig:accuracy_by_satellites} illustrates the results of the computed accuracy of target object geolocation when UAVs' GPS quality is good and excellent. In addition, the results of the experiment enable us to surround the POI with an uncertainty circle, providing the user with a clear understanding of the POI's true location.


\subsection{Usability Analysis of \dronear}
\label{sec:usability_study}
\subsubsection{Study Participants}
We recruited six participants (five men and one woman) from our network. Four of them had experience in operating UAVs or developing UAV applications. The other two were researchers in the Software Engineering domain.  All participants' ages ranged between 18 and 34. Four of them had prior knowledge of AR technologies.

\subsubsection{Study Method}

To support our user study, we recruited three volunteers to help us enact the scene of a shooting incident. Two volunteers and the first author acted as injured victims in the park, whilst one played the role of the shooter  dropping a toy gun on the ground whilst running through a park. The UAV flew autonomously in mission mode to survey the crime scene, capturing videos for use in our study. Figure \ref{fig:crime_scene} depicts the enacted scene including the location of injured victims, the toy gun - dropped by the suspect (evidence), the suspect's last known location, and his probable vehicle. As part of our study, we selected the shooting event because it was easier to enact in the real world than other incidents such as river searches or fire surveillance. 

We used the recorded aerial video stream of the artificial shooting event as described above to analyze the usability of our system. In preparation for the user study, we seeded the AR world space by using \dronear's annotation feature to add the gun as a POI so that it appeared in UI's presented to users in our study.

During our study session we first obtained participants' consent for the study, and then provided an overview of \system to familiarize them with the kinds of use-cases it would support. Second, we demonstrated \dronear's interface and asked participants to test out the annotation tool by marking a random object in an aerial video stream. Once they acknowledged confidence in using the interface, we assigned a task of monitoring the video stream of our enacted crime scene activity to: (i) acknowledge any information shared by other rescue team members or UAVs that might help them to develop an understanding of the crime scene, and (ii) annotate POIs in the aerial video stream with their corresponding symbols in order to share the information with other rescue team members. 

When participants were interacting with our system, we also recorded the screen and mouse clicks to identify any interaction pattern that participants specifically followed to complete the task. Soon after task completion, we asked participants to fill out a questionnaire and then conducted a brief follow-up interview to elicit further comments and suggestions for improvement.

\subsubsection{Results and Discussion}
We divided the post-task questionnaire into four sections. The first part of the questionnaire focused on evaluating \dronear's performance on six parameters (attractiveness, perspicuity, efficiency, dependability, simulation and novelty) using  a standard User Experience Quality (UEQ) scale \cite{laugwitz2008construction}. The second part  evaluated the overall system's usability under real-world conditions using System Usability Scale (SUS) \cite{bangor2008empirical}. In the third part, we asked participants to rate the usefulness of each \dronear feature on a 7 point Likert scale using the questionnaire shown in Table \ref{tab:feature_questions}, and finally, we asked participants to think aloud throughout the study and to provide us with insights for ways to improve the usability of \system.
Participants did not receive any compensation for participating in, or completing the study. We now describe our analysis in relation to each part of the questionnaire.

\begin{table}[]
    \centering
    \begin{tabular}{|c|p{0.6\columnwidth}|}
    \hline
    \textbf{\dronear Feature} & \textbf{Statements in our Survey} \\ \hline
     Pause Video Before Annotation & The ability to pause a video and annotate it with objects was very useful to me. \\ \hline
     Object Selection & The Object Selection Panel at the bottom was useful to pick which objects I wanted to annotate.   \\ \hline
     Supports Scene Spatial Understanding &The Gun icon gave me a good idea of where it's positioned in relation to other Vehicles in the Scene.   \\ \hline
     GeoLocation Persistence & It was useful that the system automatically stored my annotations on the video and shared them with other members of the rescue team. I do not have to share it verbally with others over the radio. \\ \hline
     POI Visualization (shared by others) & I would probably have difficulty finding the gun in the video if the gun icon did not appear automatically. \\ \hline
     Maintain Consistency across teams & I think it is better to see the evidence icons in the video stream rather than hearing where it was found over the phone or radio. \\ \hline
    
    \end{tabular}
    
    \caption{Participants rated their level of agreement with the above statements regarding the usefulness of \dronear}
    \label{tab:feature_questions}
\end{table}

\begin{itemize}
    \item \textbf{User Experience}: Figure \ref{fig:phase_1_results} (Left) provides UEQ mean scores across six dimensions. The standard deviation and confidence interval for each scale are shown. In general, the participants had a positive experience, and reported that they found \dronear to be an innovative solution for sharing information during UAV based emergency response missions, with a mean score of 2.33 for \emph{Novelty}. One participant mentioned ``\emph{I think that the system would be very useful, and it was definitely helpful being able to see the icons set up by both the teams on the ground and in charge of the aerial view. Overall, I would say that this is a very valuable tool}''. UEQ results also indicate potential for improvements in \emph{Dependability} and \emph{Efficiency} aspects of the interface. Therefore, the first author of the paper reviewed participants' feedback comments to identify improvement opportunities. 
    
    \item \textbf{Feedback}: First, two participants suggested adding the capability to rewind the aerial video stream to annotate fast-moving POIs. One of them mentioned `` \emph{For fast moving objects such as suspect, it may be useful to add a feature to rewind a video}''. This feature would also support rescue team members in revisiting existing annotations. Another participant felt the need to be able to zoom into the video to closely analyze small objects in the scene such as a gun and mentioned ``\emph{there can also be a zoom feature for the video}''. Third, two participants reported difficulties when using the UI buttons to mark objects in the aerial video stream, saying ``\emph{sometimes it is stressful to mark the targets however just because it is sort of hard to press the buttons..., but I would like to use a keyboard to press [buttons on the UI]}''. Another participant also suggested providing ``\emph{ shortcuts to select the options like gun, suspect, play, pause, etc.}'' as ``\emph{It would make it much easier to operate}''. We anticipate that providing multiple annotation methods such as keyboard shortcuts and drag-and-drop markers would improve efficiency and support diverse user interaction preferences.
    
    In the context of enhancing the situational awareness of the operators monitoring the aerial video stream, one participant suggested overlaying arrows on the edge of the aerial video stream pointing towards the POIs in the affected area. The participant said ``\emph{if there is nothing on the screen, like the times there was just grass, there could be arrows on the edges of the screen. These arrows would be pointing in the direction of where something has been marked. These arrows could be labeled with what has been marked}''. By doing so, users would be able to gain an understanding of the situation that extends beyond what is visible in the aerial video stream. As an example, if a police officer on the ground identifies critical evidence and transmits its geolocation to \system, the arrows pointing towards the evidence will benefit the rescue team in two ways. First, the direction of the arrow will allow the remote operator to understand the location of the evidence in relation to the current location of the UAV. Second, the operator may choose to use the arrows to guide the UAV towards the evidence for a closer aerial inspection and marking annotations.

\begin{figure}
    \centering
    \includegraphics[width=\columnwidth]{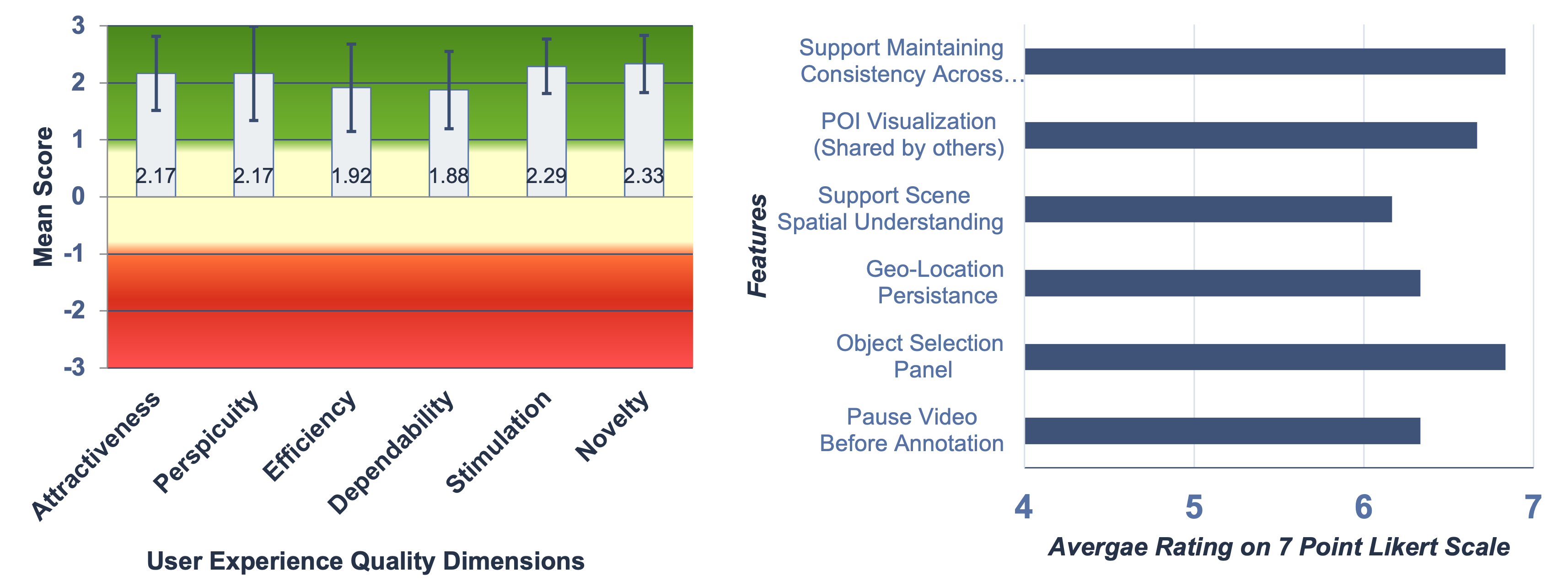}
    \caption{User Study Results : (Left) \dronear User Experience Quality Results, (Right) \dronear Features Usefulness }
    \label{fig:phase_1_results}
\end{figure}
    
    \item \textbf{Usability} : Based on the participants' responses to the statements described in Table \ref{tab:feature_questions}, we summarize the usefulness of each \dronear feature in Figure \ref{fig:phase_1_results} (Right). Participants strongly agreed that visualizing the POIs shared by other rescue team members as overlays on the aerial video stream is useful and directly supports awareness of the situation across multiple teams. When multiple POIs were available in a single frame, the POI visualization also improved the participant's spatial understanding of the scene. Participants also found other features very useful such as persistence of information, ability to pause the video stream before annotating it, and object selection panel at the bottom for picking objects. However, as the scenario gets more complex in the real world, putting many labelled buttons at the bottom of the screen might clutter the interface.  In such case, we believe replacing labels with icons on the buttons at the bottom might be more useful for providing quick access to many annotation objects.  
    Finally, we compute an SUS score of 94.6, which, based on the adjective rating scale \cite{bangor2009determining} can be defined as \textit{Excellent} (where 25 represents the worst imaginable, and 100 represents the best imaginable).

\end{itemize}

Overall, the study participants found our system to be useful and reported positive experiences with it. The participants provided valuable UX recommendations to consider for improving UI design of \dronear.

\section{Stakeholders Perspectives}
\label{sec:domain_experts}
To gain a deeper understanding of emergency rescue teams' communication needs and to understand how \system can be leveraged and improved to serve those needs during the course of investigations, we reached police officers from a local law enforcement agency and conducted a one-hour focus group session \cite{morgan1996focus}. During this session we (i) explored the challenges they face when communicating with each other during emergency response situations such as mass shootings, (ii) asked whether \system could alleviate some of those challenges, if so, how? (iii) gathered feedback regarding how \system could be utilized in real-world emergency circumstances, and finally (iv) discussed opportunities for conducting field exercises in the future.

\subsection{Participants}
Invitations to the focus group were sent via emails and resulted in participation of five police officers (four Male and one Female). Table \ref{tab:participants_details} describes the participant's backgrounds.
\begin{table}[htbp]
    \centering
        
    \begin{tabular}{|c|c|c|c|} \hline
         \textbf{Participant Id}& \textbf{Current Designation}&\textbf{Years of Service} & \textbf{Experience with UAVs}\\ \hline
         
P1& Police Chief &17&None\\  \hline
P2&Deputy Chief of Police&27&1-3 years \\ \hline
P3&Assistant Police Chief&40&Occasional at Incident Scenes \\ \hline
P4&Special Events Program Manager&15& 0-1 Years\\ \hline
P5&Deputy Chief Safety Services&34& None\\ \hline

    \end{tabular}
\caption{Details of participants in our focus group session}
    \label{tab:participants_details}
\end{table}

\subsection{Procedure}
We opted to conduct a focus group, because they have been shown to increase the likelihood of discovering a wide variety of attitudes, knowledge, and experiences in one session. The first and second authors facilitated the focus group, and the second author also took notes during the session for analysis. Participants of the focus group were asked to discuss openly their experiences, attitudes, and opinions. 

We opened the session with a presentation of the overarching goal of leveraging UAVs as partners in emergency response missions such as river search and rescue and fire surveillance, and city surveillance, and then drew on news articles detailing how police have previously used drones in emergency situations to inspire their thinking about UAV use. We then asked them about their current communication practices and challenges during emergencies. 
 
In the second half of the session, we introduced the participants to \system as a tool for sharing  geolocation information. We also presented videos of each of the  \dronear, \humanar, and \missionplan interfaces created from the enacted shooting scene described in Section \ref{sec:eval}. Finally, we asked the participants to complete a five-minute survey questionnaire in order to capture their views and opinions about the system. The entire focus group session was completed in one hour.

\subsection{Findings}

\subsubsection{\textbf{Challenges in current communication system}}
During the session, the police officers confirmed our previous understanding that they communicated primarily verbally via radio. However, they also explained that they had tablets in each police car which they used to exchange written communication. 
Most participants agreed that radio communication suffers from excessive chatter and that superfluous information is frequently broadcast when multiple people talk or communicate simultaneously. Moreover, \textit{P5} pointed out that, in large-scale operations, this can create a chaotic situation, resulting in confusion among rescue personnel. \textit{P1} also stated that radio communications between dispatchers and callers can be frustrating due to the time-sensitive nature of the situation. While discussing how geolocation information is communicated over the radio and how police officers interpret it, we learned that officers frequently refer to landmarks, roads, buildings, and other known entities in their environment. In this context, \textit{P1} shared that it can be difficult identifying locations when several buildings have similar names. 

Their current communication infrastructure lacked the capability to visualize information in any form.  However, \dronear and \humanar interfaces could alleviate this problem by enabling team members to use POI markers to share and interpret geolocation information during an emergency response. 

\subsubsection{\textbf{ geolocation importance}}
Police officers reported that geolocation assists them in many ways -- for example, by narrowing down their search or containment areas. They also agreed that tracking down a suspect becomes easier if the suspect's past geolocations are known. In the context of a mass shooting crime scene, police officers determined that the location of victims, suspects, evidence, and danger areas are of critical importance. Further, they suggested reducing the types of icons to use for POI markers to include only the essential types. \system can be customized to allow tagging aerial video stream with a limited set of markers appropriate to each scenario type.

\subsubsection{\textbf{Design}}
The police officers also suggested the following ways to improve the design of \system, in order to address the police officers' real-world needs.

\begin{itemize}
    \item \textit{Suspect Tracking}: Police officers found it helpful to mark suspects in the aerial video stream in order to determine their geolocation. They expressed interest in the UAV to tracking the suspect and for \dronear to move the POI marker according to the suspects' last known position. Integrating a tracking feature into  \dronear requires extending its implementation to continuously update the POI symbols in the AR space to reflect the last known geolocation of the suspect. 
    
    \item \textit{Textual Annotations}: In addition to using markers or symbols, police officers suggested placing textual annotations in the scene via \dronear. They believed that on-scene rescue teams would benefit from a detailed textual description of specific POIs, such as the model of a gun and the type of bullet found on the crime scene.
    
    \item \textit{Annotation Modifications}: Police officers are often required to confirm the presence of evidence during an investigation. Therefore, our participants suggested that on-scene responders should be able to move, delete, or even add new POIs using \Humanar.  This would allow a police office to augment, correct, or create information that would then be available to all team memebers, including UAVs, via \dronear and \Missionplan. As a result, all rescue teams would be able to maintain consistent awareness of the situation.

    \item \textit{System Integration}: Finally, our participants also believed that integrating \system with other situational awareness tools such as Fuses\cite{fususONE38:online} which  assists police officers in developing situational awareness of the scene by providing real-time video footage from CCTV cameras. The police chief (\textit{P1}) demonstrated Fuses to us in order to explain how it helps them in criminal investigations, and discussed the possibility of integrating video imagery from CCTV cameras via Fuses with the more flexible, mobile UAV views via \dronear interfaces.

\end{itemize}

\subsubsection{\textbf{More Applications}} Police officers identified several additional use cases. The two that they were most enthralled by were related to major sporting events on the campus and included traffic planning and searching for lost children.  Other ideas supported parking services and escort services.

\subsubsection{Police Officer's Perception of \system}

\begin{table}[htbp]
    \centering
    \begin{tabular}{|c|p{40em}|}  
    \hline
    \textbf{Qs} & \textbf{Statement} \\
	\hline
	Q1 & I believe the system's ability to geolocate POIs could benefit the police officers in carrying out their investigations \\
	\hline
	Q2 & I believe the visualizing the POIs using AR Glasses can assist on-scene officers in quickly assessing the situations \\
	\hline
	Q3 & The system supports police officers in sharing and understanding the geolocation information during emergency response \\
	\hline
	Q4 & The system could be a valuable addition to our existing communication system \\
	\hline

\end{tabular}
    \caption{Detailed description of the statements included in the survey questionnaire for police officers}
	\label{tab:survey_questions}
\end{table}

\begin{figure}[htbp]
    \centering
    \includegraphics[width=\columnwidth]{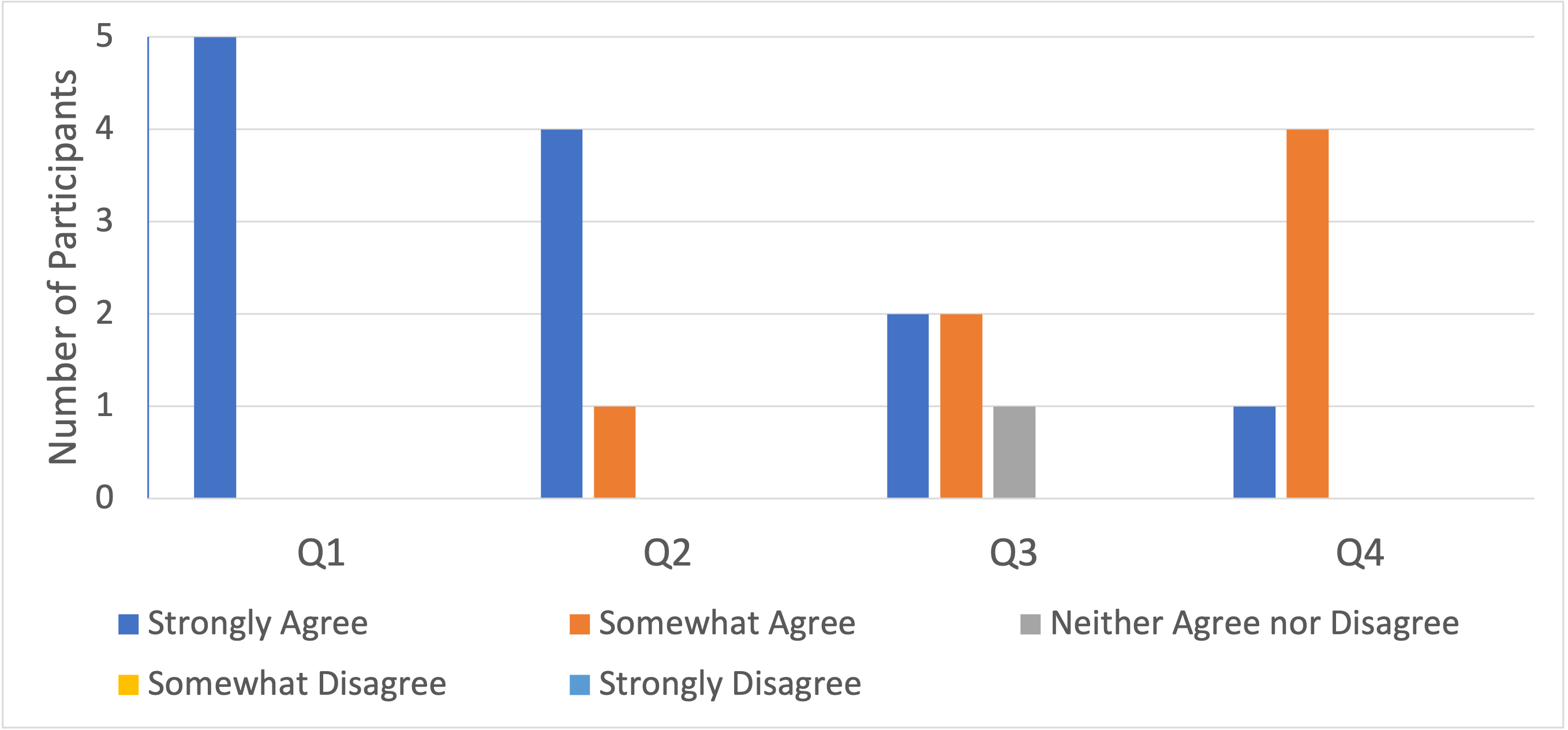}
    \caption{Officers' agreement with the survey questionnaire described in Table \ref{tab:survey_questions} on a five point Likert scale.}
    \label{fig:police_survey_questions}
\end{figure}

We developed a short questionnaire to obtain feedback from police officers regarding their perception of the usability of the \system during crime investigations, and asked police officers to rate their agreement or disagreement with the statements in Table \ref{tab:survey_questions} on a 5-point Likert scale. The survey results are shown in Figure \ref{fig:police_survey_questions}.  All police officers strongly agreed that the ability of the \system to compute the geolocation would be useful to carry out their investigations. Four strongly agreed, and one moderately agreed that visualizing the critical POIs using AR glasses would allow officers to quickly access the situation. This response suggests that \humanar could improve the response time of on Scene emergency rescue teams. Further, police officers believed that the overall system could support teams in sharing and understanding the geolocation information (two strongly agreed, two somewhat agreed, and one neural) and could be a complementary addition to the existing communication system (one strongly agreed, and four somewhat agreed). These responses suggest the positive feedback from the end-users of \system and support our design concept to improve the communication among rescue teams during emergency response. 
\section{Discussion and Future Work}
\label{sec:disscusion}
Our research proposed a novel system and multiple interfaces for improving information exchange among rescue teams and UAVs during emergency response. Based on our feasibility analysis, usability analysis, and feedback from domain experts, we found that \system presents a pragmatic approach to alleviate the social and organization challenges of communication enabling multiple rescue teams and UAVs to share and visualize scene information. Our findings in this study inform several insights to design the AR-based communication system for the next-generation UAV-Driven emergency response.  
 
First, domain experts emphasized the importance of geolocations of POIs in the scene. Therefore, it is crucial for Location-Based AR communication systems to determine the geolocations of the target objects in the scene in order to develop a feasible solution. According to our geolocation accuracy analysis discussed in Section \ref{sec:accuracy_analysis}, \system is able to calculate the geolocation of the target objects in an aerial video stream within a range of 2 meters under ideal conditions, making it a potentially valuable asset for emergency response teams. However, as our solution depends heavily on the GPS accuracy of the UAVs' position in the air, adding more sensory information to improve the UAVs' GPS signal quality would enhance the geolocation computation of the target object. For instance, leveraging Real-Time Kinetics (RTK) base stations over GPS, which provide centimeters level accuracy \cite{stott2020ground}, could enhance the target geolocation computation in urban areas where UAVs' GPS accuracy tends to deteriorate. 

Second, in our usability study discussed in Section \ref{sec:usability_study}, we noticed that four out of six users preferred to pause the aerial video before annotating it in order to ensure precision of annotations, whilst some users requested the ability to rewind the live video feed in order to replay parts of it, so that they could have sufficient time to annotate everything in the scene. In the context of this observation, two factors should be taken into consideration when designing \dronear. First, if users are allowed to pause and rewind the live video feed during a fast-paced emergency response mission, they will not have time to analyze the incoming video frames for the duration of the pause or for the time required to complete the annotation. Second, preventing the users from pausing the video or rewinding it would prevent them from annotating the video if there are many points of interest within one frame. Thus, in order to assure that users are able to analyze every video frame as well as pause and rewind the live aerial video feed, user interface design choices should be explored. A possible solution could be to use the aerial video buffer to show the aerial video at a fast rate in order to catchup on the time taken for annotating.  Thus, future research could examine options for addressing such trade-offs in UX design of \dronear. Further, from the perspective of user experience design, we believe that the users should be able to choose whether they wish to skip the video frames or view it at a higher frame rate in such a scenario. 

Finally, our study with domain experts discussed in Section \ref{sec:domain_experts} provided us with end-user requirements that we should consider while designing AR supported emergency rescue tools. The demonstration of the \system prototype resulted in an agreement to conduct future field exercises with them to evaluate \dronear in a real world setting.

\section{Limitations}
\label{sec:limitations}
Our work is subject to several limitations. We have illustrated that the system can provide communication support during the enactment of a shooting incident;  however, further analysis is needed for more complex emergency response scenarios. In addition, the system must undergo additional field-testing with police officers and other stakeholders to ensure its robustness. We considered several alternative designs to evaluate the accuracy of our system in computing the geographical location of the target object, including the possibility of collecting and analyzing the data in real-time. We plan to study the robustness of our system during future field tests with stakeholders. However, our approach of using pre-recorded videos of the scene and the geolocation of the target objects allowed us to control the UAVs' GPS signal strength and to conduct a thorough accuracy analysis of the geolocated POIs. The analysis can be extended to high-speed UAVs flying at much higher altitudes.  
Another threat is related to the experience of participants in the focus group. While all police officers have served for more than 15 years in the department, their experience in the use of UAVs was limited, and two police officers had no experience in working alongside UAVs. Therefore, we anticipate more use-cases and design improvements for our UAV-based emergency response system. 
\section{Conclusion}
\label{sec:conclusion}
In this paper, we have shared our vision for using AR during emergency response to make geolocation information more accessible in order to support informed and quick decision-making. We introduced \System, a location-based Augmented Reality platform that consists of three different user interfaces: \Dronear, \Humanar, and \Missionplan. These interfaces allow rescue personnel and UAVs to share and visualize geolocated POIs using AR. We conducted multiple studies, including a focus group session with end-users of our platform, to study the feasibility and usability of \system in the real world. We found that \system is capable of calculating the geolocation of POIs in the scene within a range of 2 meters under ideal conditions. In addition, participants in our usability study indicated that \dronear assists rescue team members in maintaining a consistent awareness of the scene. In our focus group discussion, police officers also expressed their belief that our system has the potential to benefit law enforcement agencies in the course of investigations, and could be a valuable addition to their existing communication systems. Finally, we also identified design implications for building future collaborative UAV driven AR systems. 

\section{Acknowledgements}
\label{sec:ack}
Double blinded

\bibliographystyle{ACM-Reference-Format}
\bibliography{main}

\end{document}